\documentclass[11pt,a4paper]{article}

\usepackage{amsmath,amssymb,wrapfig}
\usepackage{epsfig,graphicx}
\usepackage[font={small},labelfont=bf]{caption}
\usepackage{subfigure}
\usepackage{graphicx}
\usepackage{rotating}
\usepackage{cancel}
\usepackage{bm}
\usepackage{color}
\usepackage{comment}
\usepackage{cite}
\usepackage{psfrag}

\renewcommand\[{\left[}

\newcommand{\exclude}[1]{}
\newcommand{\etmiss}{E^{\text{miss}}_{T}}

\def\beq{\begin{equation}}
\def\eeq{\end{equation}}

\begin{document}
\numberwithin{equation}{section}
\title{
\vspace{2.5cm} 
\Large{\textbf{Getting Stuck!}}\\[0.2cm]
\large{\textbf{Using Monosignatures to Test Highly Ionizing Particles
\vspace{0.5cm}}}}

\author{Christoph Englert$^{1}$ and Joerg Jaeckel$^{2}$\\[2ex]
\small{\em $^1$SUPA, School of Physics and Astronomy, University of
  Glasgow,}\\
\small{\em Glasgow G12 8QQ, United Kingdom}\\[0.8ex]
\small{\em $^2$Institut f\"ur Theoretische Physik, Universit\"at Heidelberg,} \\
\small{\em Philosophenweg 16, 69120 Heidelberg, Germany}\\[0.5ex] }

\date{}
\maketitle

\begin{abstract}
\noindent In this paper we argue that monojet and monophoton searches can be a sensitive test of very highly ionizing particles such as particles with charges $\gtrsim 150e$ and more generally particles that do not reach the outer parts of the detector. 8 TeV monojet data from the CMS experiment excludes such objects with masses in the range $\lesssim 650~{\text{GeV}}$ and charges $\gtrsim 100e$. This nicely complements searches for highly ionizing objects at ALICE, ATLAS, CMS and LHCb. Expected improvements in these channels will extend the sensitivity range to $m\lesssim 750~{\text{GeV}}$. This search strategy can directly be generalized to other particles that strongly interact with the detector material, such as e.g. magnetic monopoles.
\end{abstract}

\thispagestyle{empty}

\newpage


\section{Introduction}
Monojet and monophoton searches are a popular tool to search for particles interacting so weakly with the detector that they do not leave an observable trace. In particular they have become one of the main avenues to search for dark matter particles at colliders, see e.g.~\cite{atlasphoton,cmsphoton,atlasdilepton,cmslepton,atlaslepton,cmsjets}.

At the very opposite extreme one can imagine particles that interact so strongly with matter that they are stopped before they reach essential parts of the detector, such as the calorimeters, which is required for triggering and eventual detection. In some cases they may be stopped even before they leave the beam pipe, such that there is no chance that they can be directly detected with the usual multi-purpose detectors. Examples of candidate particles interacting very strongly with matter are particles with high electric charges\footnote{We focus on this  scenario mainly as a simple test case. Nevertheless from the theoretical point of view charges in U(1) gauge field theories can a priori be completely arbitrary. However, one should admit that most embeddings into more complete models favor charges not much larger than unity. That said, large hierarchies of charges have recently been discussed in the context of solutions to the hierarchy problem, e.g.~\cite{Kaplan:2015fuy}.} or magnetic monopoles (perhaps even with multiple magnetic charges)~\cite{Dirac:1931kp,Dirac:1948um}. This shortcoming of the Large Hadron Collider's multi-purpose experiments is part of the motivation of the MoEDAL experiment~\cite{Pinfold:2009oia}, which has published first results in Ref.~\cite{MoEDAL:2016jlb}.

The main goal of this note is to point out that monojet and monophoton searches at the multi-purpose detectors are also a powerful tool to search for such super-strongly interacting particles, which nicely complement and extend the sensitivity reach of these LHC experiments.
\begin{wrapfigure}[9]{r}{0.49\textwidth}	
\hfill\parbox{0.46\textwidth}{
\centering
\includegraphics[width=0.24\textwidth]{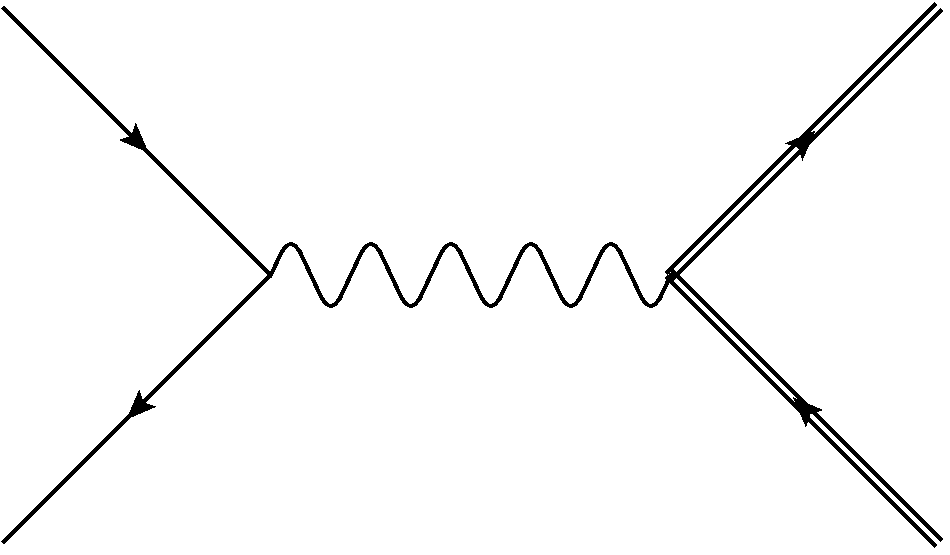}
\caption{Drell-Yan production of a highly charged particle (indicated by a double line).}
\label{drell}}	
\end{wrapfigure}
 The idea is quite straightforward: if the produced particles are so strongly interacting that they are stopped before they leave a recognizable trace in the detector they are just as ``invisible'' as particles that are very weakly interacting. This sets the limitation in general searches for highly ionizing particles (HIPs) in Drell-Yan-like production~\cite{DeRoeck:2011aa}, Fig.~\ref{drell}. However, if these objects have a large mass in the 100 GeV range, this mass scale will induce a shower signature to the full hadronic final state via coherent initial and/or final state radiation. Since this is a purely kinematics-driven phenomenon based on QCD factorisation, at least the appearance of initial state radiation should not be dependent on the particular production mechanism of the HIP and will occur even if production is intrinsically non-perturbative, which can be expected for extended objects such as monopoles. This way, even if we have no triggerable signature (identically to very weakly interacting particles), there is a potential signature from the recoil of the HIP against additional Standard Model emission, leading to monophoton, mono-$Z$ and monojet signatures.
 
We will demonstrate in this paper that highly charged particles can be constrained at the LHC employing ``standard'' dark matter searches. This can be generalised
to searches for other highly ionising particles such as magnetic monopoles. We will argue that monojet and monophoton searches are highly complementary to the on-going search efforts at the LHC, but also complementary to MoEDAL, leading to large increase of sensitivity to these scenarios.

This brief note is structured as follows. In the following two sections we start with the concrete example of particles with large electric charges. We recall how they are stopped in Sec.~\ref{stopping} and what this implies for searching them with monojet and monophoton searches. The actual limits are presented in Sec.~\ref{limits}. To put this into perspective we discuss some potential issues with highly charged particles and their description in Sec.~\ref{schwinger}. 
In Sec.~\ref{general} we then discuss more general aspects applicable to wider classes of particles. We end with concluding words in Sec.~\ref{conclusions}.

\section{Producing and stopping particles with high electric charges}
\label{stopping}
To demonstrate the general idea of searching for very strongly interacting particles with monojet and monophoton searches we consider the concrete example of highly charged particles, more concretely highly charged fermions.
But even at this point we would like to stress that the approach more generally applies to particles that have a significant chance of being stopped before they reach an essential part of the detector.

\subsection{Productions with monojet and monophoton}\label{production}
\begin{figure}[!b]
\centering
\includegraphics[width=0.24\textwidth]{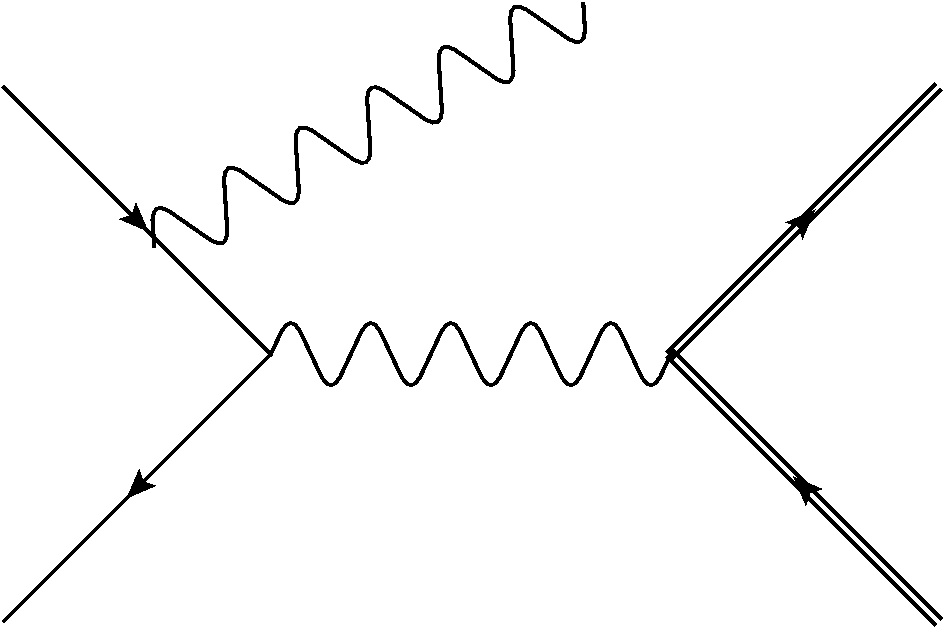}
\hspace*{1cm}
\includegraphics[width=0.24\textwidth]{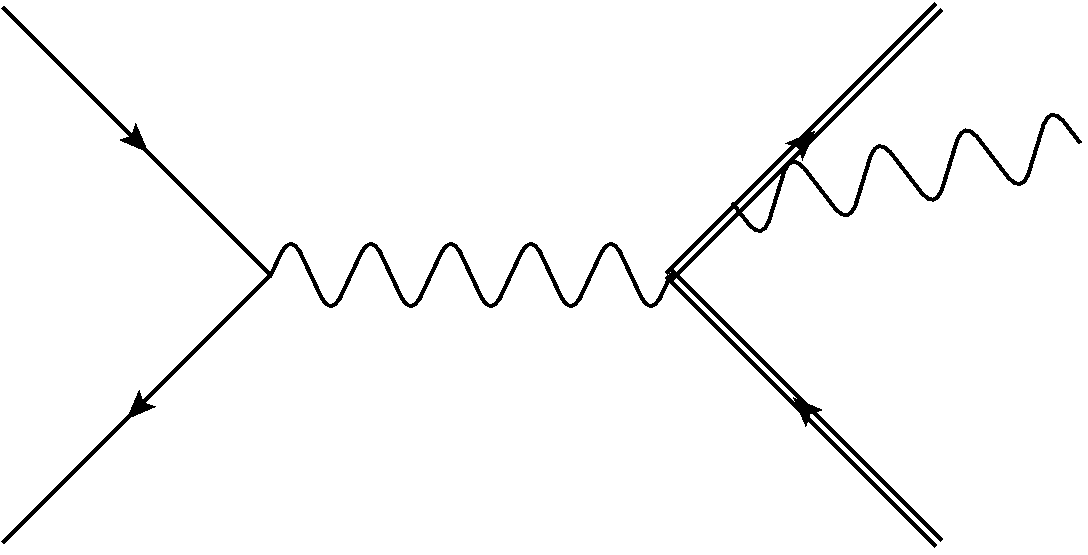}
\caption{Representative diagrams for Drell-Yan production leading to a monophoton signature. The photon can be radiated off the initial but also the final state particle, see text.}
\label{monophoton}	
\end{figure}

The simplest production mechanism of highly charged particles is Drell-Yan (cf.~Fig.~\ref{drell}), which is typically employed to provide estimates for collider searches (see~\cite{Pinfold:2009oia,DeRoeck:2011aa,Chatrchyan:2013oca,Aad:2013gva,Khachatryan:2014mea,Aad:2014gfa,ATLAS:2014fka,Khachatryan:2015jha,Aad:2015oga,Aad:2015rba,Aad:2015kta,CMS:2015kdx,Aaboud:2016dgf,Aaboud:2016uth}). However, if the 
%
\begin{wrapfigure}[11]{r}{0.49\textwidth}
\hfill\parbox{0.46\textwidth}{
\centering
\includegraphics[width=0.24\textwidth]{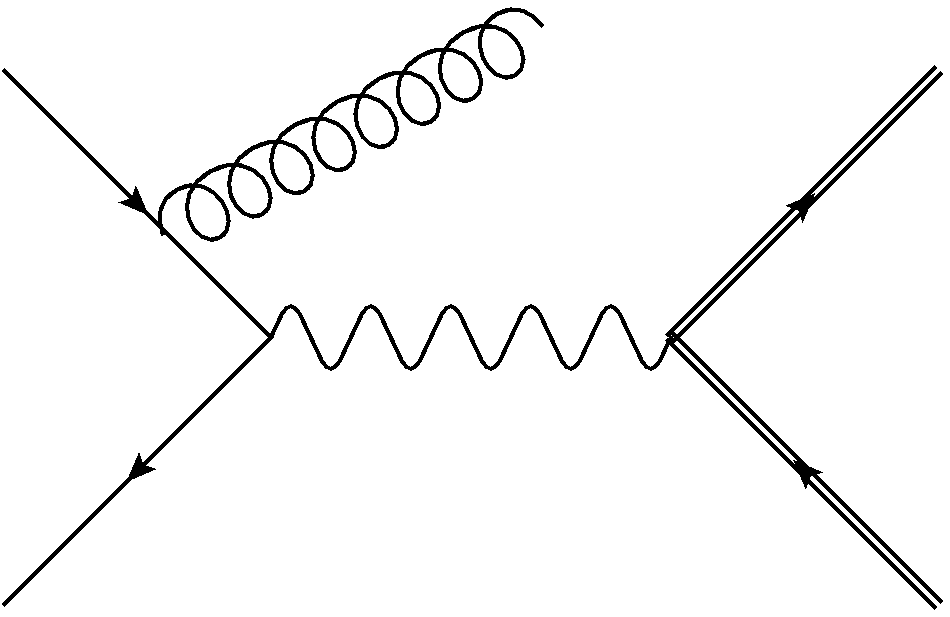}
\caption{Representative Feynman diagram of Drell-Yan production of a highly charged particle (double line) in association with the production of a single gluon from initial state radiation, eventually giving rise to a monojet signature.}
\label{monojet}	}
\end{wrapfigure}
%
%
particles are stopped they alone will not produce a detectable signal. We will therefore consider 
monojet and monophoton processes, examples of which are depicted Figs.~\ref{monojet}~and~\ref{monophoton}.
\subsection{Stopping highly charged particles}
\label{sec:stop}
An essential ingredient for our search strategy is that the produced highly charged particles are stopped early enough. Let us now estimate when and how this happens.

To do so we follow~\cite{DeRoeck:2011aa} and use the Bethe-Bloch formula for the energy loss of charged particles with velocity $\beta$, gamma-factor $\gamma$ and charge $z$ (in units of $e$) in materials,
\begin{equation}
-\frac{dE}{dx}=K\frac{Z}{A}\frac{z^2}{\beta^2}\left[\log\left(\frac{2m_{e}\beta^2\gamma^2}{I}\right)-\beta^2\right]\,.
\end{equation}
Here $x$ is the amount of material per area traversed, effectively,
\begin{equation}
x=\rho \ell
\end{equation}
where $\ell$ is the distance travelled in the medium of density $\rho$.
 
In the Bethe-Bloch equation we also have the constant,
\begin{equation}
K=0.307\,{\text{MeV}}\,{\text{g}}^{-1}\,{\text{cm}}^{2}\,,
\end{equation}
and we need to specify the material properties.
$Z$ and $A$ are the nuclear charge and nucleon number of the atoms of the medium. Finally $I$ is the mean excitation energy in the medium, which is documented in the PDG review~\cite{Agashe:2014kda} for the two relevant materials we consider,
\begin{eqnarray}
~^{9}\text{Be}&:& A=9, \,\,\,\,\, Z=4,\,\,\,\,\,\,\, I=63.7\,{\text{eV}}\qquad {\text{beam pipe}}\,,
\\\nonumber
~^{28}\text{Si}&:& A=28,\,\,Z=14,\,\,\,\,I=137\,{\text{eV}}\qquad \,\,{\text{inner tracker}}\,.
\end{eqnarray}

To determine whether the particle gets stuck we need to integrate the Bethe-Bloch equation up to the point where the highly charged particle has lost all its kinetic energy, i.e. $\gamma=1$. Using $E=\gamma m$ one can easily scale out the mass and charge dependence from the Bethe-Bloch equation. The distance travelled in a given (fixed) material is then given by,
\begin{equation}
\label{stoppi}
x_{\text{stop}}=\left(\frac{100}{z}\right)^2\left(\frac{m}{1\,{\text{TeV}}}\right) f_{\text{mat}}(\gamma)\,\frac{g}{{\text{cm}}^2}\,.
\end{equation}
The function $f_{\text{mat}}$ only depends on the material and the initial gamma-factor of the highly charged particle. 
Integrating the Bethe-Bloch equation for different values of $\gamma$ we find the following fitting functions,
\begin{equation}
f_{\text{Be}}(\gamma)=\bigg\{\begin{array}{lcc}
0.25& {\text{for}} & \gamma\lesssim 1.05\\
29.7766 - 69.2787 \gamma + 39.2006 \gamma^2& {\text{for}} & 1.05\lesssim \gamma\lesssim 1.5\\
-72.5575 + 57.5741 \gamma - 0.159835 \gamma^2& {\text{for}} & 1.5\lesssim \gamma\lesssim 50\\
\end{array}
\end{equation}
and 
\begin{equation}
f_{\text{Si}}(\gamma)=\bigg\{\begin{array}{lcc}
0.26& {\text{for}} & \gamma\lesssim 1.05\\
28.6498 - 67.4852 \gamma + 38.5221 \gamma^2& {\text{for}} & 1.05\lesssim \gamma\lesssim 1.5\\
-69.0793 + 55.7173 \gamma - 0.167437 \gamma^2& {\text{for}} & 1.5\lesssim \gamma\lesssim 50
\end{array}\,.
\end{equation}
We will use these functions directly in our Monte Carlo simulation, they are accurate within about 15\%.

We can now determine whether a particle of given mass, charge, momentum and direction will be stopped in the beam pipe or in the inner tracker. As in~\cite{DeRoeck:2011aa} we use a ``thickness'' of the beam pipe and inner tracker of
\begin{eqnarray}
{\text{beam pipe}}: &&d_{\text{bp}}=0.148\,\frac{\text{g}}{{\rm cm}^2}\quad {\text{(Beryllium)}}
\\\nonumber
{\text{inner tracker}}: &&d_{\text{it}}=9.6\,\frac{\text{g}}{{\rm cm}^2}\quad\quad\, {\text{(Silicon)}}.
\end{eqnarray}

For a given energy, mass and pseudo-rapidity the minimal charge required to stop the particle is then given by
\begin{equation}
\label{stopcut}
Q_{\text{min}}(E,\eta)=100\sqrt{f\left(\frac{E}{m}+1\right)\left(\frac{m}{\text{TeV}}\right)\frac{1}{x(\eta)}}
\end{equation}
with
\begin{equation}
x(\eta)=\frac{d}{\sin(\theta)}=\frac{d}{\sin\left(2\arctan[\exp[-\eta]]\right)}\,,
\end{equation}
where $d$ is either $d_{\text{bp}}$ for the beam pipe or $d_{\text{it}}$ for the inner tracker. The angular/pseudo-rapidity dependence arises because particles at small angles have to traverse more material before leaving the device in question.

For particles stopping in the inner tracker we would in principle also have to take into account the combined effects of the beam pipe material and that of the inner tracker. However since the thickness of the beam pipe is more than an order of magnitude smaller than that of the inner tracker, we have neglected the effects of the beam pipe when considering the stopping in the inner tracker thereby effectively loosing a small amount of stopping power which makes this approximation conservative.

\section{Limits from Monojet and Monophoton searches}
\label{limits}
\subsection{Monojet searches}
As a first example let us now consider the limits from the CMS monojet search~\cite{cmsjets}. We focus on the 8 TeV results, as the recent 13 TeV analyses employ model specific assumptions~\cite{CMS-PAS-EXO-16-010,CMS-PAS-EXO-16-038,CMS-PAS-EXO-16-039,CMS-PAS-EXO-16-037}, making a direct comparison less reliable. The size of the data set is also not important for the qualitative impact of monojets on the discussed scenarios; we will extrapolate the 8 TeV limits to the 13 TeV run 2 expectation as well as to the end of the high luminosity phase.

To obtain these limits we have simulated the monojet process (cf. Fig.~\ref{monojet}) and accepted only those events that are stopped in the respective parts of the detector, either the beam pipe or the inner tracker.
Effectively this corresponds to a $Q$ and $\eta$-dependent cut on the energy of the produced particles.
The explicit value of this cut can be obtained by solving Eq.~\eqref{stopcut} for the energy as a function of $Q$ and $\eta$. We require that both highly charged particles are stopped sufficiently early.

As already discussed in the previous section we consider two options where the particles can be stopped in order to be ``invisible'' and the signal to be counted in the monojet search (or the monophoton search below).
It is clear that particles stopped in the beam pipe will not leave any detectable trace in the detector\footnote{Except maybe some $\delta$-radiation originating from the stopping process.}. The situation is not so clear in the inner tracker. However, as the particle is stopped before it reaches the calorimeter, it does not leave an observable cluster there that could be associated to the stopped track. In any case, the large misbalance in transverse momentum for these events will leave these events triggered.
Therefore, those events can be understood from a monojet search~\cite{private-communication}. To be on the safe side we will perform our analysis for both cases (stopping in the beam pipe and in the inner tracker, respectively) and will reserve the interesting question of how detectable, yet non-standard signatures can be used to tighten constraints on HIPs to future work. Some analyses in these directions are in the process of being prepared or already performed. For the former category there are efforts to enhance the trigger sensitivity to so-called ``track stubs''~(see e.g. \cite{Warren:1234900,Hoff:2012nz}), which provide a serious challenge for the high luminosity phase of the LHC that we discuss below. Sensitivity to such objects will also allow to put stringent constraints on the models that we discuss in this paper.

Analyses along these lines, which are already performed are exotics searches via disappearing tracks, see e.g. \cite{Aad:2013yna,CMS:2014gxa}. These are typically model-dependent but could be recast to the scenarios discussed in this work. Massively charged objects such as monopoles provide another motivation to continue these analyses even if the targeted model space in, say, supersymmetric scenarios with long-lived particles becomes excluded.

In this paper we focus on the simplest strategy to constrain such a scenario: looking at the monojet signals (the jet allowing it to be triggered) it should be possible to do a search for extra tracks/energy deposition in the inner tracker caused by the stopping of the strongly interacting particle. This is a clearly different signature from the Standard Model background events and should allow for a significant reduction in the background.

\begin{figure}[!t]
\centering
\includegraphics[width=0.47\textwidth]{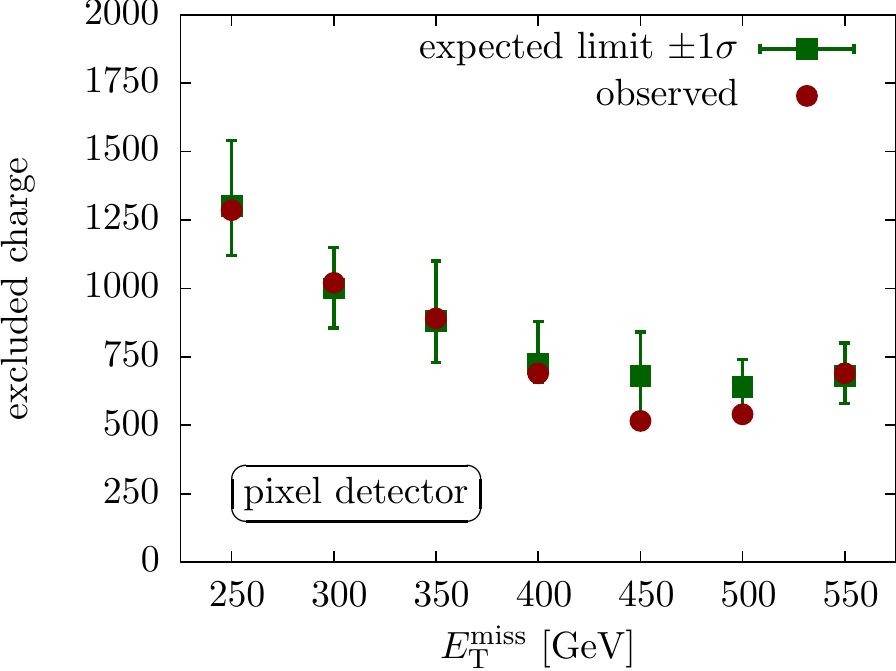}
\hfill
\includegraphics[width=0.47\textwidth]{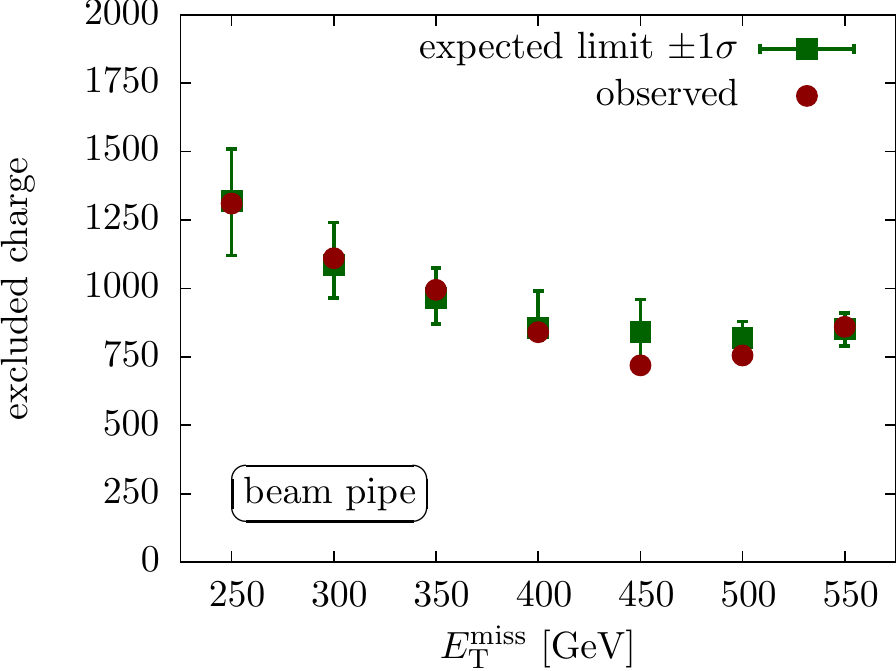}
\caption{Limits on the maximal charge (in units of $e$) of a particle for the different $\etmiss$ bins considered in the CMS search~\cite{cmsjets}. Particles are required to be stuck in the inner tracker (left panel) or the beam pipe (right panel).
For both examples we have chosen a particle mass of $m=1000~\text{GeV}$.}
\label{binning}	
\end{figure}

Let us now try to obtain some first limits, through interpreting the analysis of~\cite{cmsjets} in the present context. CMS define signal regions from inclusive $\etmiss$ bins. The details of the analysis (which reports a model-independent limit at 8 TeV) can be inferred from~\cite{cmsjets}, but we summarise
\begin{wrapfigure}[22]{l}{0.49\textwidth}
\parbox{0.48\textwidth}{\includegraphics[width=0.47\textwidth]{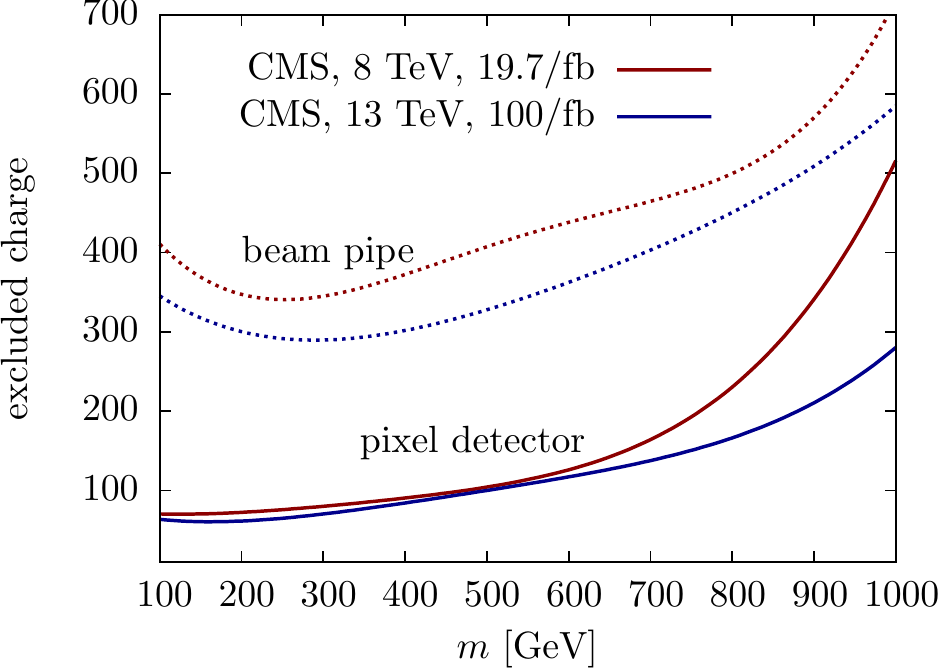}
\caption{Limits on highly charged particles (in units of $e$) from the monojet search. The production process is assumed to be Drell-Yan plus jet.  The dotted lines labelled ``beam pipe'' only take into account signal events where both produced highly charged particles are stopped in the beam pipe. The solid ``pixel detector'' lines require particles to be stopped in the inner tracker.}\label{monojetlimits}}
\end{wrapfigure}
the key selection criteria here for completeness. 

The CMS analysis is based on clustering jets with the anti-$k_T$ algorithm~\cite{Cacciari:2008gp} with a resolution parameter $D=0.5$. The leading jet in the event
needs obey the following transverse momentum and rapidity requirements
\begin{equation}
  p_{T,j_1}>110~\text{GeV}, |\eta_{j_1}| < 2.4\,,
\end{equation}
and an additional jet
\begin{equation}
  p_{T,j_2}>30~\text{GeV}, |\eta_{j_2}| < 4.5
\end{equation}
is only allowed if it is consistent with a monojet signature
\begin{equation}
  \Delta\Phi(j_1,j_2) < 2.5\,.
\end{equation}
Events with more than two jets in the 
$p_T>30~\text{GeV}$ and $|\eta_j|<4.5$ regions are vetoed as well as events with isolated leptons if $p_{T,\ell}>10~\text{GeV}$. CMS define ``isolation'' by
requiring that the total hadronic energy deposit in a cone of size $\Delta R\equiv \sqrt{\Delta\Phi^2+\Delta \eta^2}< 0.4$
around the lepton candidate is smaller than 20\% of the candidate's $p_T$. 
The analysis focuses on seven inclusive search regions defined by the amount of observed missing energy.

In Fig.~\ref{binning} we have considered each of these selections separately and rephrased the CMS constraints as a constraint on the maximal allowed charge of the particle for the setup described in Sec.~\ref{sec:stop}. This demonstrates that the bins with higher $\etmiss$ are more sensitive for the case of a $m=1~\text{TeV}$ state as indicated in Fig.~\ref{monojet}. The reason for this is that the high mass scale induced by the HIP leads to a signal that is clustered at relatively hard jet transverse momentum distribution $p_{T,j}$. Given that the contributing backgrounds are steeply falling distributions as a function of the $\etmiss \sim p_{T,j}$, high $\etmiss$ selections provide the tightest constraints for large HIP masses.

Taking the bins from the CMS search into account and using the CLs method~\cite{Read:2002hq} to extrapolate the results of the 8 TeV analysis to 13 TeV, we obtain the  limits shown in Fig.~\ref{monojetlimits}. 

\subsection{Monophoton search}
In analogy to the monojet search we can also use monophoton searches. As already mentioned in Sec.~\ref{production} there is one crucial difference to the monojet signal. The photon can also arise from final state radiation (cf. right hand part of Fig.~\ref{monophoton}). In principle this could be advantageous because it significantly increases the signal. In this situation there are two electromagnetic interactions with the highly charged particle. Therefore, the final state radiation contribution to the cross section scales as,
\begin{equation}
\sigma({\text{monophoton}})\sim Q^4.
\end{equation}
This is much larger than the\footnote{Of course this also indicates a breakdown of perturbation theory. We will briefly discuss this issue in the next section.}
\begin{equation}
\sigma(\text{monojet})\sim Q^2\, ,
\end{equation}
in the case of monojet (cf. Fig.~\ref{monojet}) or that of only initial state radiation (cf. left hand side of Fig.~\ref{monophoton}).

For comparability we again choose to recast an analysis by CMS~\cite{cmsphoton}.

Again, anti-$k_T$ jets with $D=0.5$ are used
and isolated photons are reconstructed through a similar isolation strategy as leptons:
The energy deposit in a cone $\Delta R
=\sqrt{\Delta\Phi^2+\Delta \eta^2} < 0.3$ has to be smaller than 5\% of the photon candidate's energy.
We require at least one isolated photon with $E_T(\gamma)>145~\text{GeV}$ in $|\eta|\leq 1.44$. 
Events with more than a single jet with $p_T>30~\text{GeV}$ and light leptons
(isolation is based on a hadronic energy deposit in the vicinity of
$\Delta R <0.3$ by less than 20\% of the candidate's $p_T$) with
$p_T>10~\text{GeV}$ are vetoed if they are well-separated from the photon
by $\Delta R>0.5$. In the last step the analysis focusses on a large missing transverse
energy ${{E}}_T^{\text{miss}}>140~\text{GeV}$, whose direction in the transverse plane needs to be well-separated (and back-to-back) to the photon $\Delta \Phi(E_T^{\text{miss}},\gamma)>2$.

\begin{figure}[!t]
\centering
\includegraphics[width=0.47\textwidth]{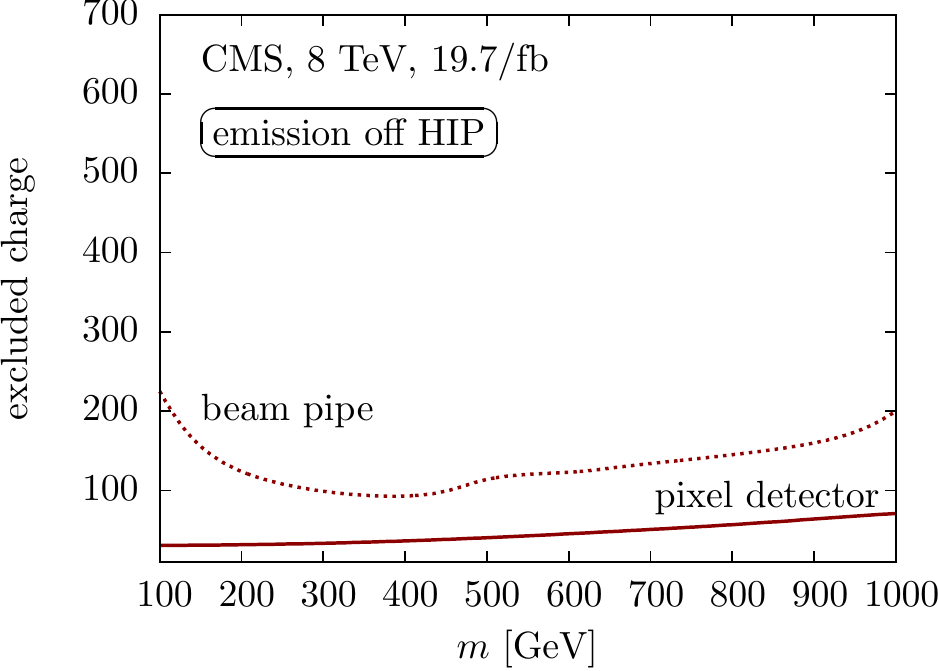}
\hfill
\includegraphics[width=0.47\textwidth]{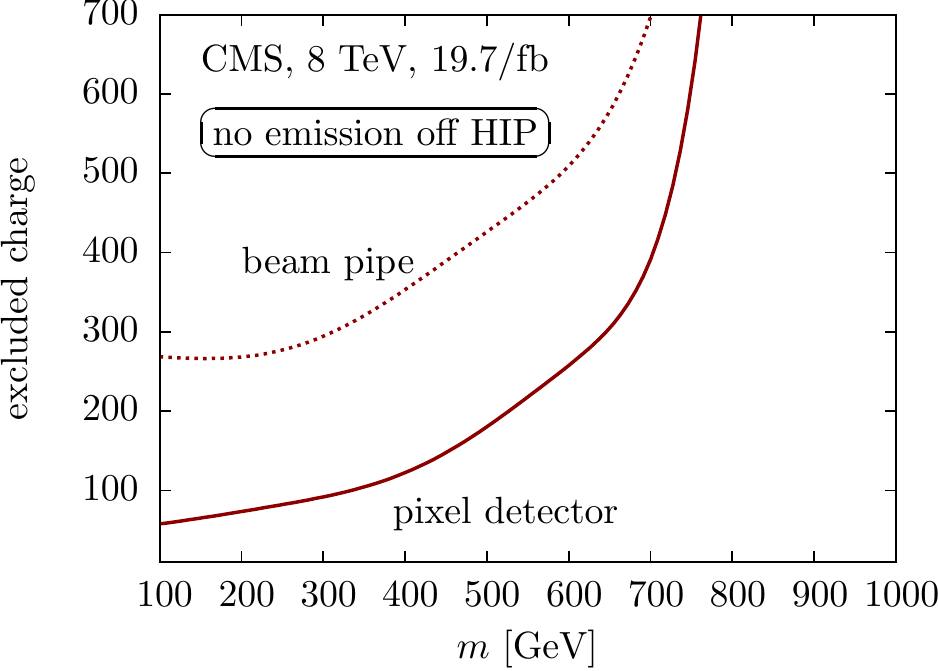}
\caption{Limits on highly charged particles (in units of $e$) from the monophoton search~\cite{cmsphoton}. As in Fig.~\ref{monojetlimits}
the dotted line only considers events stopped in the beam pipe, whereas the solid ones also allows for events stopped in the inner tracker the full monophoton signal, including both initial and final state radiation. In the right panel we only take into account initial state radiation.}
\label{monophotonlimits}	
\end{figure}

The dependence of the production cross section on the charge for the different emission scenarios is reflected in the significantly better limits shown as the red lines in Fig.~\ref{monophotonlimits} (left panel) compared to the monojet analysis. 

Similar conclusions directly generalise to mono-$Z$ analyses, which, however are less sensitive compared to the monophoton case, and we therefore do not discuss them in detail.

However, as we will briefly discuss in the following section, we are in a deeply non-perturbative regime for highly charged particles. Therefore a simple perturbative treatment of the final state radiation (or perhaps even the production process itself) may be dubious. We therefore also show in Fig.~\ref{monophotonlimits} (right panel) the limits taking into account only initial state radiation such that the cross section is,
\begin{equation}
\sigma({\text{initial state radiation only, monophoton}})\sim Q^2\,.
\end{equation}
We regard this as a reasonably conservative estimate.

\section{Conceptual and technical issues with highly charged particles}
\label{schwinger}
As already indicated in the previous sections there are some problematic issues with highly charged particles.
While we have ignored them in our general analysis, let us nevertheless briefly mention some of them in this section. In the next section we then discuss how the monophoton searches can be interpreted in a more general and perhaps slightly less model-dependent way.

\subsubsection*{Non-perturbativity}
The first problem with highly charged particles is that we quickly enter a regime where QED is non-perturbative.
The relevant expansion parameter is,
\begin{equation}
\alpha Q^2\sim 10^{-2}\, Q^2.
\end{equation}
Even if we generously assume perturbativity up to $\alpha Q^2\sim 4\pi$ we still have,
\begin{equation}
\alpha Q^2\sim 4\pi,\quad{\text{for}}\quad Q\sim 40\,,
\end{equation}
severely limiting the range of perturbation theory.

\subsubsection*{Landau pole}
Related to the question of non-perturbativity is that the Landau pole of QED will be very close to the {\it elementary} particle mass threshold and therefore within the region probed by the LHC. 

For large charges we can neglect the contribution of the SM particles. The Landau pole is then at
\begin{equation}
\Lambda=m\exp\left(\frac{3\pi}{\alpha(m)Q^2}\right)\sim \exp(1)\quad {\text{for}}\quad Q\gtrsim 35\,.
\end{equation}
Again for sufficiently large charges $Q\gtrsim 35$ the Landau pole is less than a factor of $e$ above the particle mass.

\subsubsection*{Schwinger pair production}
The two problems discussed above affect the consistency of the theory. However, in principle there
is also a practical consideration that may limit the stopping power of the high charges: at very high charges
the field is self-shielding due to electron-positron pair production via the Schwinger mechanism~\cite{Schwinger:1951nm}. Let us briefly also estimate this effect. 

Schwinger pair production is a non-perturbative effect analog to tunneling. Its rate per volume is given by
\begin{equation}
\frac{d\Gamma}{dV}\sim m^4_{e}\left(\frac{E}{E_{c}}\right)^2\exp\left(-\frac{\pi E_c}{E}\right)\,,
\end{equation}
where $E$ is the electrical field, $m_{e}$ the electron mass and we have used the critical field for the production of electron-positron pairs,
\begin{equation}
E_{c}=\frac{m^{2}_{e}}{e}.
\end{equation}
Pair production becomes fast when the critical field is exceeded in a volume larger than the Compton volume of the electron $V\sim 1/m^{3}_{e}$. 

The electric field of a highly charged particle is,
\begin{equation}
E_{Q}\sim \frac{Qe}{4\pi r^2}.
\end{equation}
If the field exceeds the critical field strength at a distance of two Compton length the condition for rapid pair production is certainly fulfilled in a volume of $\sim 8m^{-3}_{e}$. This is the case when
\begin{equation}
\frac{Qe^2}{16\pi}=\frac{\alpha Q}{4}\gtrsim 1\quad {\text{for}}\quad Q\gtrsim 500.
\end{equation}
Requiring rapid pair production only in a smaller volume $\sim m^{-3}_{e}$ the bound can be tightened by a factor of $4$ to $Q\sim 100$. Comparing with the results from the previous section we see that self-shielding from pair production might be a significant effect.

Finally, we note that in particular the Schwinger pair production mechanism is special to the case of highly charged particles. For example for magnetic monopoles this effect is not expected.
\section{Monojet and Monophoton searches for general stoppable particles}\label{general}
The objections discussed in the previous section are to a large degree specific to highly charged point-like particles. Non-elementary or extended objects such as magnetic monopoles may not suffer from all of these concerns.

Indeed our strategy to search for particles strongly interacting with the detector medium is more general. As long as the particle is produced from quarks it is likely that at least initial state radiation (gluons, photons or $Z$s) can be produced. Events where the produced new particles are stopped will then contribute to the monojet or monophoton signal.

In general it is difficult to determine model-independent limits. As discussed in Sect.~\ref{stopping} the fraction of events where the particle is stopped sufficiently early depends in general on the energy and angular distribution. Moreover, it will also depend on the energy dependence of the stopping process. 
That said, let us nevertheless take one step in the direction of a more general limit. Assuming that all these dependencies are the same as in the Drell-Yan + jet or Drell-Yan + photon process, we can at least allow for an arbitrary production cross section, e.g. the particles could additionally be produced from decays of other particles or there could be a form factor suppressing the production. 

\begin{figure}[!t]
\centering
\includegraphics[height=0.4\textwidth]{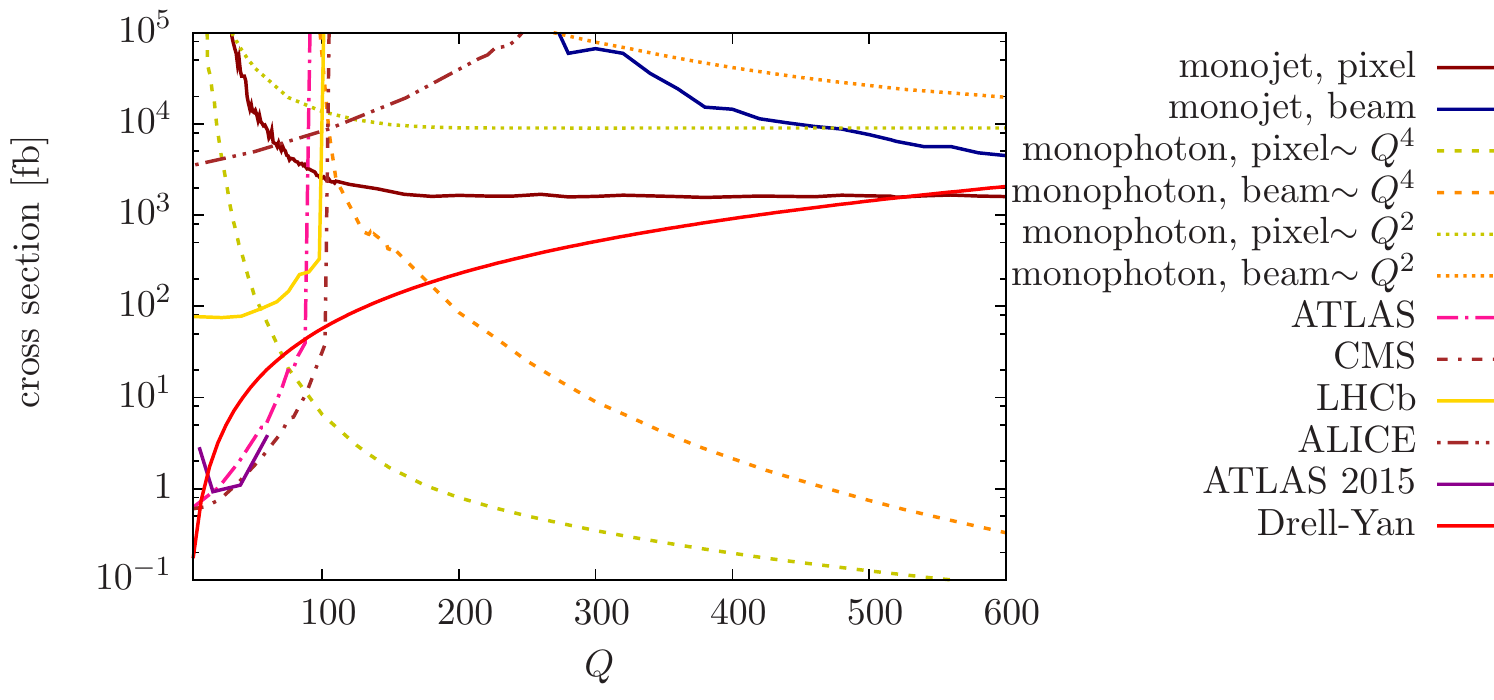}
\caption{
Limits on the production cross section from the monojet and monophoton searches as described in the text. For the angular and energy distributions we have assumed the same as for Drell-Yan + jet. Similarly we have assumed that the stopping has the same energy distribution as for a highly charged particle of mass $1000~{\text{GeV}}$. We also include the projections by ATLAS, CMS, LHCb and ALICE according to Ref.~\cite{DeRoeck:2011aa} (which is obtained for 7 TeV collisions) as well as 2015 ATLAS constraint~\cite{Aad:2015kta}.}
\label{modelindependent}	
\end{figure}

\begin{figure}[!t]
\centering
\includegraphics[height=0.4\textwidth]{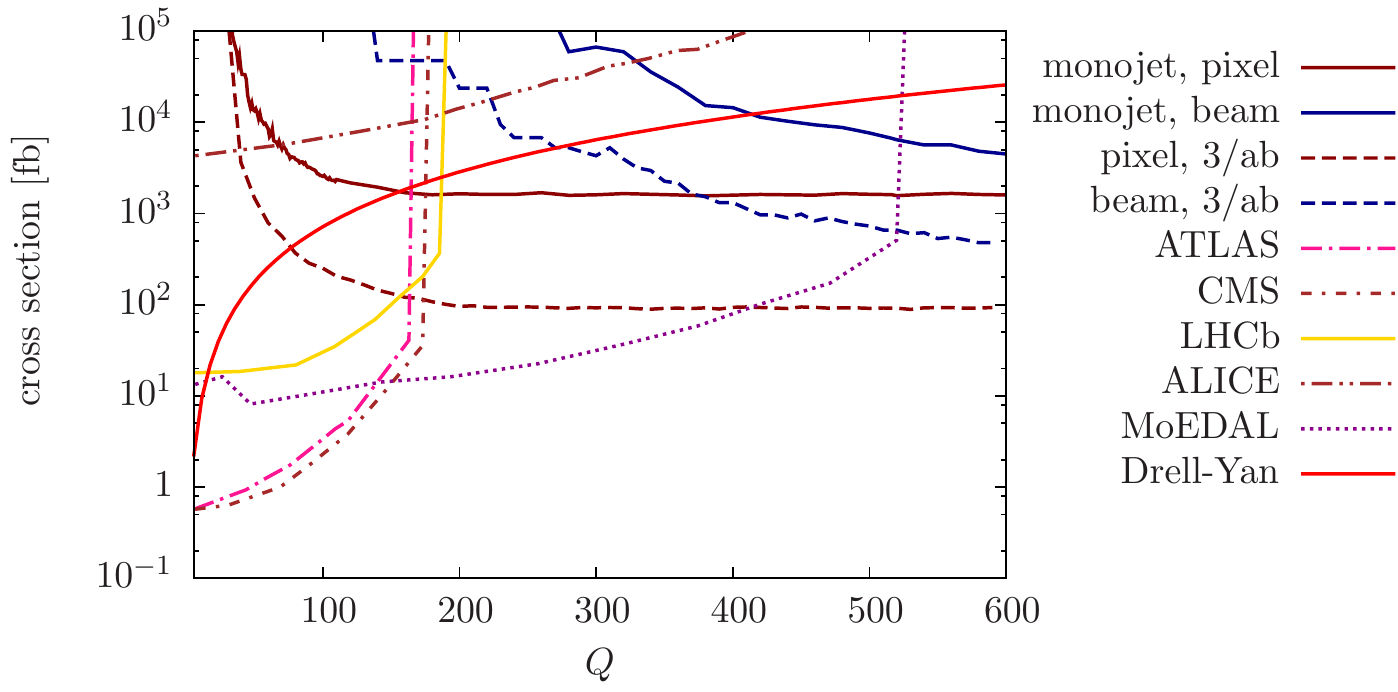}
\caption{Same as Fig.~\ref{modelindependent} with 14 TeV projections from Ref.~\cite{DeRoeck:2011aa}. We also include monojet projections for the LHC high-luminosity run with $3000~\text{fb}^{-1}$ with conservative 13 TeV centre-of-mass energy, reducing the all systematics with the square root of the luminosity compared to Fig.~\ref{modelindependent}.}
\label{modelindependent2}	
\end{figure}
Other searches by ATLAS, CMS, LHCb, ALICE and MoEDAL are directly based on the Drell-Yan production cross section. To facilitate the comparison with our results we show in Fig.~\ref{modelindependent} our cross section limit expectations  re-scaled to that of the Drell-Yan production without jets, i.e. we multiply by
\begin{equation}
\frac{\sigma({\text{Drell-Yan}})}{\sigma(\text{Drell-Yan +jet})}\,.
\end{equation}
This also allows for a comparison between the sensitivity of the monojet and monophoton searches in Fig.~\ref{modelindependent}, where we also show extrapolations to 13 TeV data taking for the monojet case, which is in general more sensitive when HIP emission is not included (and when the perturbative expansion is more reliable).

\bigskip

While the limits presented in this section are far from model-independent, they at least give an indication
of the reach of the monojet and monophoton searches for particles that can be stopped in the material of the detector. While we expect details such as the angular distribution to be different, we also expect that in general particles with higher energy/momentum are less likely to be stopped. In consequence only those new particles that are produced with relatively small energy/momentum will contribute to the monojet/monophoton signal - a feature at least qualitatively represented by our Drell-Yan example.

\section{Concluding Remarks}
\label{conclusions}
In this note we have argued that monojet and monophoton searches are powerful tools in searches for new particles that interact so strongly with the detector material that they are stopped before they reach parts of the detector which are essential for recording these events.
If all of the new particles are stopped in the beam pipe there will be no signature in the actual detector\footnote{Except maybe from cascades caused by the stopping process, e.g. $\delta$-electrons.}.
If there is additional initial and/or final state radiation the event will look like a monophoton or monojet final state. For the purposes of monojet and monophoton searches the same is likely to hold if the new particles are stopped in the tracker. The detector response to the new particles lacks the features (signals in the calorimeters or the muon detectors) that will allow the experiments to identify them as electrons, muons or jets. Consequently, the corresponding event is likely to be counted as part of a monosignature analysis.
Yet, on closer inspection of the detector response such events will be very different from the SM background monojet/monophoton events, thereby opening the chance for significantly improved searches.

Here we have focussed our calculations on the concrete example of highly charged particles. The very high charges considered may raise questions about perturbativity, calculational methods and even the viability of the model (nevertheless we essentially use the same technical calculation of cross sections etc. as used in related searches\footnote{However, pointing out one caveat is in order. The potential shielding caused by Schwinger pair production would probably increase the signal in searches looking for highly ionizing tracks, while decreasing the monosignatures. However, this is largely specific to the case of highly charged particles.}).  Importantly, however, the general strategy proposed here should be more widely applicable, in particular also to the case of magnetic monopoles. Our analysis in Sect.~\ref{general} is only a first step in this direction. To obtain a wider coverage and (at least partially) address the issues of perturbativity a sensible next step could be to study a variety of reasonable production distributions combined with different models for the stopping process in the material. The results from this could then be turned into limits on the production cross section similar to the discussion in Sect.~\ref{general}.
Nevertheless, the example of highly charged particles constrained by the recent CMS searches for monojets and monophotons demonstrates that our proposed search strategy has significant power complementary to existing search strategies at ATLAS, CMS, LHCb, ALICE as well as dedicated detectors such as MoEDAL, motivating further studies.

\subsection*{Acknowledgements}
We thank Hans-Christian Schultz-Coulon and Shahram Rahatlou for helpful discussions. We also thank the organisers of the 2016 Patras Workshop for providing the environment where this work was initiated.
J.J. gratefully acknowledges support by the DFG TR33 ``The Dark Universe'' as well as the European UnionÕs Horizon 2020 research and innovation programme under the Marie Sklodowska-Curie grant agreement Numbers 674896 and 690575.

\providecommand{\href}[2]{#2}\begingroup\raggedright\endgroup

\end{document}